\documentclass[aps,twocolumn,groupedaddress,showpacs]{revtex4}
\usepackage{graphicx}
\def\sigmav{{\mbox{\boldmath{$\sigma$}}}}
\begin{document}
\bibliographystyle{apsrev}
%

\title{UNUSUAL SYMMETRIES IN THE KUGEL-KHOMSKII HAMILTONIAN}
\author{A. B. Harris}
\address{Department of Physics and Astronomy, University of
Pennsylvania, Philadelphia, PA 19104}
\author{Taner Yildirim}
\address{NIST Center for Neutron Research,
National Institute of Standards and Technology, Gaithersburg, MD
20899}
\author{Amnon Aharony, Ora Entin-Wohlman, and I. Ya. Korenblit}
\address{School of Physics and Astronomy, Raymond and Beverly Sackler
Faculty of Exact Sciences,  Tel Aviv University, Tel Aviv 69978,
Israel }

\date{\today}
\begin{abstract}

The Kugel-Khomskii Hamiltonian for cubic titanates describes spin
and orbital superexchange interactions between $d^1$ ions having
three-fold degenerate $t_{2g}$ orbitals. Since orbitals do not
couple along ``inactive" axes, perpendicular to the orbital
planes, the total number of electrons in $|\alpha \rangle$
orbitals in any such plane and the corresponding total spin are
both conserved.  A Mermin-Wagner construction shows that there is
no long-range spin ordering at nonzero temperatures. Inclusion of
spin-orbit coupling allows such ordering, but even then the
excitation spectrum is gapless due to a continuous symmetry. Thus,
the observed order and gap require more symmetry breaking terms.
\end{abstract}

%
%


\maketitle


High temperature superconductivity\cite{B} and colossal
magnetoresistance\cite{CMR} sparked much recent interest in the
magnetic properties of transition metal oxides, particularly those
with orbital degeneracy\cite{KKrev,TN}. In many transition metal
oxides, the $d$ electrons are localized due to the
large on-site Coulomb interaction, $U$. Assuming a simple Hubbard
model, with a typical nearest neighbor (nn) hopping energy $t$,
the low energy behavior can be described by an effective
superexchange model, which involves only nn spin and orbital
coupling, with energies of order $\epsilon=t^2/U$. In cubic oxide
perovskites, the crystal field of the surrounding oxygen octahedra
splits the $d$-orbitals into a two-fold degenerate $e_{g}$ and a
three-fold degenerate $t_{2g}$ manifold. In most cases, these
degeneracies are further lifted by a cooperative Jahn-Teller (JT)
distortion\cite{KKrev}, and the low energy physics is well
described by an effective superexchange spin-only model
\cite{ANDERSON,TY1,TY2}. However, some cubic perovskites, such as
the titanates (RTiO$_{3}$, where R= La, Y, etc), have only a small
JT distortion\cite{JTdist}, in spite of the orbital
degeneracy\cite{noJT}. This distortion, as well as the small
orthorhombic deviation from the cubic symmetry, were not even
observed in Refs. \onlinecite{LTO,YTO}. Since these distortions
are small, several theoretical papers chose to ignore them and
{\it assume} cubic symmetry. The corresponding cubic model has
been taken as the ``minimal" model needed to explain the physics
in these materials. In the present Letter we show that although
this model is of great theoretical interest, it is insufficient to
explain the experiments.

For the cubic titanates, there is one $d$ electron in the $t_{2g}$
degenerate manifold, which contains the wavefunctions $|X \rangle
\equiv d_{yz}$, $|Y \rangle \equiv d_{xz}$, and $|Z \rangle \equiv
d_{xy}$. The large degeneracy of the resulting ground states,
which involve {\em both} the spin and the orbital degrees of
freedom\cite{KKrev,TN,KK}, may then yield rich phase diagrams,
with exotic types of order, involving a strong interplay between
the spin and orbital sectors (e.g. \onlinecite{TN,LTO,YTO}),
justifying the broad theoretical interest in this cubic limit. As
we show, the corresponding superexchange Hamiltonian (hereafter
called the cubic Kugel-Khomskii (KK) model\cite{KK}) contains
several interesting hidden symmetries. In addition, our analysis
shows that the KK Hamiltonian {\it cannot} yield some of the
predictions which were claimed in the literature to follow from
it. In particular, it has been suggested \cite{GK1} that the KK
Hamiltonian gives rise to an ordered isotropic spin phase at
non-zero temperatures, and that an energy gap in the spin
excitations can be caused by spin-orbit interactions \cite{GK2}.
We use the symmetries of the KK Hamiltonian to show that both of
these predictions cannot hold. The observed long range order and
finite gap \cite{LTO} {\it must} therefore be based on more
complicated Hamiltonians, which go beyond the scope of this paper.

\begin{figure}
\includegraphics[scale=0.4]{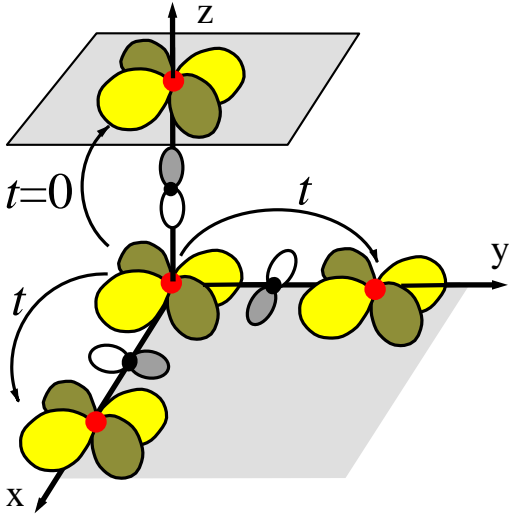}
\bigskip
\bigskip
\bigskip
\caption{A schematic
view of the $|Z\rangle = d_{xy}$ orbitals and the (indirect)
hopping parameter $t$ via intermediate oxygen p-orbitals.
Positive (negative) regions of wavefunctions are represented by
dark (light) lobes. Symmetry forbids indirect hopping
along the z-axis for an electron in the Z-orbital.}
\label{DXY}
\end{figure}

As can be seen from Fig. \ref{DXY}, cubic symmetry implies no
hopping (via oxygen $p$-states) among orbitals of type $|\alpha
\rangle$ along the $\alpha$-axis. Neglecting the direct Ti-Ti
hopping, KK called this axis the ``inactive axis'' for $\alpha$
orbitals. This statement forms the basis for the remarkable
symmetry properties of the KK Hamiltonian reported in this Letter.
Apart from constant terms, the perturbative expansion of the
Hubbard Hamiltonian with hopping of this type, to order
$\epsilon$, yields the cubic KK Hamiltonian, ${\cal H} = {\cal
H}_x + {\cal H}_y + {\cal H}_z$, where
\begin{eqnarray}
{\cal H}_\alpha = \epsilon \sum_{\langle ij \rangle \in \alpha}
\sum_{\beta , \gamma \not= \alpha} \sum_{\sigma , \eta} c_{i
,\beta , \sigma}^\dagger c_{i ,\gamma , \eta} c_{j ,\gamma ,
\eta}^\dagger c_{j ,\beta , \sigma}. \label{NOW}
\end{eqnarray}
and $\langle ij \rangle \in \alpha$ denotes a nn bond along the
$\alpha$-axis. Here, $c_{i, \beta , \sigma}^\dagger$ creates an
electron at site $i$ in a $\beta$ orbital with spin $\sigma$, and
one assumes that there is exactly one electron on each site, i.e.
$\sum_\alpha n_{i\alpha}=1$, with $n_{i\alpha} \equiv \sum_\sigma
c_{i ,\alpha , \sigma}^\dagger c_{i ,\alpha , \sigma}$.

For some purposes, it is convenient to separate the spin and
orbital degrees of freedom\cite{KK}. Defining the spin of an
electron at site $i$ as ${\bf S}_i$, one has
\begin{eqnarray}
{\cal H}_\alpha = (\epsilon/2)\sum_{\langle ij \rangle \in \alpha}
\left( 1 + 4 {\bf S}_i \cdot {\bf S}_j \right) J_{ij}^\alpha  \ ,
\label{KKEQ} \end{eqnarray}
\begin{eqnarray}
J_{ij}^z = n_{ix} n_{jx} + n_{iy}n_{jy} + a_i^\dagger b_i
b_j^\dagger a_j + b_i^\dagger a_i a_j^\dagger b_j.
 \label{jij}
\end{eqnarray}
Here, $a^\dagger_i$ and $b^\dagger_i$ create {\it spin-less}
electrons in orbitals $|X \rangle$ and $|Y \rangle$, respectively,
 and $n_{ix}=a^\dagger_i a_i$ etc.

 Both Eqs. (\ref{NOW}) and
(\ref{jij}) imply that whenever an $\alpha$-orbital electron is
destroyed, an $\alpha$-orbital electron is created on either the
same or another site. Therefore, the total number of electrons in
each orbital is a good quantum number: any eigenfunction can be
labeled by the total number of electrons {\it in each orbital}
(i.e. $N_X, N_Y,$ and $N_Z$)\cite{IHM}. Furthermore, when an
$\alpha$-orbital electron is destroyed, it is replaced by another
$\alpha$-orbital electron {\it in the same plane perpendicular to
the inactive ($\alpha$) axis.}  Thus, for the $n$'th plane
perpendicular to the $\alpha$-axis, the total number $N_{n\alpha}$
of electrons in orbital $|\alpha\rangle$ is conserved, i. e. it is
a good quantum number. For example, for a cube of eight sites, the
numbers $N_{1X}$ and $N_{2X}$, which respectively are the numbers
of X-orbital electrons in each of the two planes perpendicular to
the $x$-axis, are  conserved, and similarly for $y$ and $z$. Thus,
the states of the cube can be labeled by the six quantum numbers,
$(N_{1X},N_{2X},N_{1Y},N_{2Y},N_{1Z},N_{2Z})$.

Remarkably, there are more conserved quantities, associated with
electron spins. Defining the spin of an electron in orbital
$\alpha$ at site $i$ as ${\bf S}_{i\alpha} \equiv
\sum_{\rho\eta}c_{i ,\alpha , \rho}^\dagger
[\vec\sigmav]_{\rho\eta}c_{i ,\alpha , \eta}/2$, where $\vec
\sigmav$ represents the vector of Pauli matrices, and the total
spin of all such electrons - located in an arbitrarily chosen
plane \#$n$ perpendicular to the inactive $\alpha$ axis - as
$\vec{\cal S}_{n \alpha} \equiv \sum_{i \in n}{\bf S}_{i\alpha}$,
we next perform a uniform, but arbitrary, rotation of all the
${\bf S}_{i\alpha}$'s with $i \in n$: we introduce an arbitrary $2
\times 2$ unitary matrix ${\bf U^{(n)}}$, and write
\begin{eqnarray}
c_{i,\alpha, \sigma}^\dagger &=& \sum_\eta U^{(n)}_{\sigma , \eta}
d_{i,\alpha, \eta}^\dagger \ ,\ \ \ i \in n. \label{SYMEQ}
\end{eqnarray}
Electrons in other orbitals or in other lattice planes are not
affected by this transformation. Substitution of this
transformation into Eq. (\ref{NOW}) shows that it leaves ${\cal
H}$ invariant. As a consequence of this symmetry, if one assumes
long-range spin order, the spins associated with $\alpha$-orbitals
within any given plane can be rotated at zero cost in energy,
thereby destroying the supposed
long-range order. It also follows that $\vec{\cal S}_{n \alpha}$
commutes with ${\cal H}$, and thus both $|\vec{\cal S}_{n
\alpha}|^2$ and $[\vec{\cal S}_{n \alpha}]_z$ are good quantum
numbers for each value of $n$ or $\alpha$. These symmetries can
also be obtained from the original Hubbard model, provided one
neglects Coulomb exchange interactions.

This situation allows a {\it rigorous proof} of the nonexistence
of long-range spin order at any nonzero temperature for ${\cal H}$
of Eq. (\ref{NOW}).  Following the procedure of Mermin and Wagner
(MW)\cite{MERMIN}, we choose
\begin{eqnarray}
C &=& \hat S^+_\alpha({\bf k})=\sum_{\bf R} e^{-i {\bf k} \cdot
{\bf R}} c_{{\bf R}, \alpha , \uparrow}^\dagger
c_{{\bf R}, \alpha , \downarrow}\ , \nonumber \\
A &=& \hat S^-_\alpha({\bf k+K})=\sum_{\bf R} e^{i ({\bf k} + {\bf
K}) \cdot {\bf R}} c_{{\bf R}, \alpha , \downarrow}^\dagger
c_{{\bf R}, \alpha , \uparrow}\ ,
\end{eqnarray}
where ${\bf K}$ is the wavevector of the order we wish to discuss.
Here $a{\bf K}= (\pi , \pi , \pi)$ is most relevant. Assuming long
range  order of $\hat S^z_\alpha({\bf K}) \equiv s_{\alpha,z}$ and
a corresponding staggered field $h$ (for spins in the
$\alpha$-orbital), we end up with the MW-like bound
\begin{eqnarray}
1 \ge 2kT|s_{\alpha,z}|^2\frac{1}{N}\sum_{\bf
k}\bigl[hs_{\alpha,z}+\hat J_\alpha({\bf k})\bigr]^{-1},
\label{MW}
\end{eqnarray}
where $\hat J_\alpha({\bf k}) \propto \sum_{{\vec
\delta}\notin\alpha}(1-e^{-i{\bf k}\cdot {\vec \delta}})$ is
proportional to the ${\bf k}$-dependent parts in the Fourier
transform of the non-zero nn spin-spin interaction (with nn vector
${\vec \delta}$) in the Hamiltonian. Since spins in orbital
$\alpha$ couple only within planes perpendicular to the
$\alpha$-axis, it follows that $\hat J_\alpha({\bf k}) \propto
2\sum_{\beta \ne \alpha}[1-\cos(k_\beta a)] \approx \sum_{\beta
\ne \alpha} a^2k_\beta^2 \equiv a^2k_{\perp,\alpha}^2$, with {\it
no dispersion in the} $\alpha$-{\it direction}. For systems in $d
\le 3$ dimensions, the sum in Eq. (\ref{MW}) diverges as $h
\rightarrow 0$, implying that $s_{\alpha,z}$ must go to zero.  The
conclusion is that the KK model is at its lower critical dimension
$d_<=3$ and does not support long-range spin order at $T>0$. As we
show elsewhere, a similar proof can be formulated for the original
Hubbard model\cite{details}.

The same conclusion also follows from a renormalization group
analysis of the model at finite $T$\cite{details}. Generalizing to
$m$ orbitals and $n$-component spins, the spin free energy
functional maps onto that of the ``canonical" $nm$-component spin
problem \cite{DG}, but with a $(d-1)$-dimensional transverse
gradient term,
\begin{eqnarray}
F &=& {1 \over 2} \sum_{{\bf q}\alpha} (r+ q_{\perp,\alpha}^2)
\hat{\bf S}_\alpha({\bf q}) \cdot \hat{\bf S}_\alpha(-{\bf q})\nonumber \\
&+& \sum_ {\bf R} \biggl(u\sum_\alpha|\hat{\bf S}_\alpha({\bf
R})|^4+ v\sum_{\alpha<\beta} |\hat{\bf S}_\alpha ({\bf
R})|^2|\hat{\bf S}_\beta({\bf R})|^2\biggr) \ , \label{FF}
\end{eqnarray}
where $\hat{\bf S}_\alpha({\bf q})$ is the Fourier transform of
${\bf S}_\alpha({\bf R})$. Similar forms arise in connection with
Lifshitz-like behavior\cite{AARG}. This anisotropic gradient term
shifts {\it both} the upper and the lower critical dimensions up
by 1. For $3<d<5$ dimensions and $n>1$ it yields decoupled
$n$-component critical behavior. The free energy (\ref{FF}) also
reflects the symmetry with respect to independent rotations of the
spin ${\bf S}_\alpha$ associated with the single orbital $\alpha$.

\begin{figure}
\includegraphics[scale=0.4]{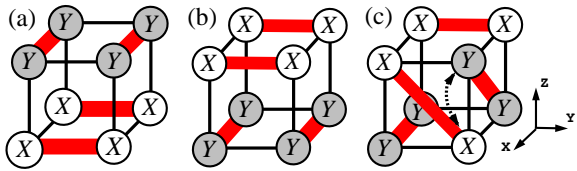}
\bigskip
\bigskip
\caption{Spin and
orbital configurations for a cube of eight sites. The thick lines
indicate singlet spin states (dimers) and the $X$ and $Y$ indicate
the orbital states of the electrons. Configurations in (a) and (b)
are the dominant ones in the ground state wavefunction. The less
dominant configuration (c) is obtained from (b) by allowing the
interchange of two ($X$ and $Y$) electrons along the $z$-axis,
retaining their membership in the spin singlets (even though their
positions have changed).} \label{hopping}
\end{figure}

These symmetries and conservation laws are very useful in the
exact numerical diagonalization of finite Ti clusters, which
indeed confirms their validity. We demonstrate this  for a cube of
8 sites (Fig. \ref{hopping}). Since the Hamiltonian commutes with
the total spin ${\bf S}=\sum_i {\bf S}_i$, the eigenstates can be
identified by the quantum numbers $S$ [where ${\bf S}^2=S(S+1)$]
and $S_z$. Since the energy does not depend on $S_z$, it suffices
to study the subspace of $70 \times 3^8=459270$ states with
$S_z=0$. A numerical analysis of the low-energy spectrum of this
huge sparse matrix yielded 3 degenerate $S=0$ ground states,
$\Psi_x$, $\Psi_y$, and $\Psi_z$, related by cyclic permutations.
$\Psi_z$ has $N_x=N_y=4$.  This information suffices to find the
manifold containing each of these ground states. Since $\Psi_z$,
for instance, is not degenerate within the manifold $N_x=N_y=4$,
it must also have the quantum numbers $N_{1X}=N_{2X}=2$,
$N_{1Y}=N_{2Y}=2$, and $N_{1Z}=N_{2Z}=0$.  A nonsymmetric choice
like $N_{1X}=3$ and $N_{2X}=1$ would be degenerate with
$N_{1X}=1,N_{2X}=3$. The lack of degeneracy also implies that the
total spin of the $N_{n,\alpha}=2$ electrons in orbital $\alpha$
in the $n$'th plane perpendicular to the $\alpha$-axis must be
${\cal S}_{n\alpha}=0$. Examples of such configurations,
containing dimers of $\alpha$-orbital electrons in the
$\alpha$-planes, are shown in Fig. \ref{hopping}. The Hamiltonian
allows an exchange of an $X$-electron with a $Y$-electron only
along the $z$-axis (the only axis along which {\it both} types can
hop). Starting from Fig. \ref{hopping}(a), and performing all such
possible exchanges, creates a manifold of 16 states (3 of which
are shown in the figure). Two other states with the same
$N_{n\alpha}$'s, but with the dimers along the $z$-axis, form
another manifold, of higher energy. Indeed, a diagonalization of
the resulting $16 \times 16$ matrix reproduced the same ground
state energy as found from the $459270 \times 459270$ matrix,
demonstrating the power and the correctness of these symmetries.
We are currently extending these numerical studies to even larger
systems (such as $N=16$ sites) to better understand the nature of
the ground state in a real system.

Since the KK Hamiltonian (\ref{NOW}) forbids long range spin order
at $T>0$, the existence of such order (as in LaTiO$_3$ \cite{LTO})
must result from some additional mechanism. Even for cubic
symmetry, such mechanisms could include the small direct Ti-Ti
hopping along the inactive axis, Coulomb exchange terms
  in the original Hubbard model, or the spin-orbit interaction. In
  the real orthorhombic titanates one must also include JT
  distortions and oxygen octahedra rotations.
  A full discussion of all these effects lies beyond the present
  paper. Since the present paper concerns mainly symmetry
  arguments,
   we concentrate here on  adding the spin-orbit interaction
   to the cubic KK Hamiltonian, where we can use such
   arguments to show the absence of a spin gap.
Specifically, the spin-orbit Hamiltonian is
\begin{eqnarray}
{\cal H}_{SO}=\lambda\sum_{i\alpha \beta \sigma
\sigma'\mu} L^\mu_{\alpha\beta}c^\dagger_{i,\alpha,\sigma}
[\sigmav_\mu]_{\sigma\sigma'}c_{i,\beta,\sigma'}\ ,
\label{SO}
\end{eqnarray}
where $L^\mu_{\alpha\beta}\equiv \langle\alpha|L^\mu|\beta\rangle$
is the orbital angular momentum matrix element. Since ${\cal
H}_{SO}$ mixes orbitals, an $\alpha$-electron can hop via ${\cal
H}_{SO}$ to orbital $\beta$, then hop to a $\beta$-orbital on a nn
along the $\alpha$-axis, and finally use ${\cal H}_{SO}$ to return
to orbital $\alpha$. This generates an effective hopping between
$\alpha$-orbitals along the ``inactive" $\alpha$-axis,
invalidating the above arguments, shifting the lower critical
dimension for total spin ordering back to $d_<=2$ and restoring
long range spin order at $d=3$. This mixing also eliminates the
independent symmetries which we found for electrons within each
orbital separately. However, as discussed below, there still
remain some global symmetries for the total spin. Based on the
signs of the leading couplings, we assume that the total spin
orders antiferromagnetically, and proceed to show that the spin
wave excitations in the ordered phase {\it must be gapless}.

Again, an exact symmetry analysis clarifies the situation. For
electrons within each of the three degenerate $t_{2g}$ orbitals
discussed here, we introduce the following canonical
transformation\cite{TY1} from spin to pseudo-spin:
\begin{eqnarray}
c^\dagger_{i , \alpha , \sigma} &=& \sum_\eta {\bf
U}^{(\alpha)}_{\sigma , \eta} d^\dagger_{i, \alpha , \eta} \ ,
\end{eqnarray}
where $U^{\alpha}= \sigmav_\alpha$ represents a different rotation
for spins in different orbitals. As discussed in Ref.
\onlinecite{TY1}, all the terms in both the original ${\cal H}$
and in ${\cal H}_{SO}$ now contain only combined operators of the
form $\sum_\mu d^\dagger_{i,\alpha,\mu}d_{i,\beta,\mu}$, with
coefficients which do not depend on the pseudo-spin indices $\mu$
(see Eq. (6) in Ref. \onlinecite{TY1}). These combined operators,
and therefore also the full Hamiltonian, are {\it rotationally
invariant in pseudo-spin space}. Said differently, the Hamiltonian
is invariant with respect to a transformation on the original spins
of the form
$\tilde c_{i,\alpha,\mu}^\dagger = \sum_{\mu'} V_{\mu , \mu'}^{(\alpha)}
c_{i,\alpha, \mu'}^\dagger$
with ${\bf V}^{(\alpha)}= \sigmav_\alpha {\bf U} \sigmav_\alpha$,
where ${\bf U}$ is an arbitrary unitary matrix.  Thus the system
possesses a continuous symmetry, but it is not the usual symmetry
with respect to rotation of the total spin. In the
antiferromagnetically ordered phase, the spin staggered moment
selects an orientation, and therefore the pseudo-spin will also
exhibit broken symmetry. Rotation of the pseudo-spin will give
rise to a manifold of zero energy states. This continuous symmetry
guarantees that we have a (probably propagating) zero-energy
hydrodynamic mode\cite{TCL}. The rigorous conclusion is then that
spin-orbit interaction permits the existence of long-range order
at nonzero temperatures, but does not cause a gap in the
elementary excitation spectrum, contrary to the assertion in Ref.
\onlinecite{GK2}. Since our argument is based on symmetry
considerations, it holds no matter what type of fluctuation is
considered, and regardless of the orbital ordering (long ranged or
liquid). In analogy with results of Refs. \onlinecite{TY1,TY2,BS},
it is probable that when Coulomb exchange interactions and/or
canting of the Ti--O--Ti bonds are included, spin-orbit
interactions would lead to an energy gap in the excitation spectrum.

In conclusion,  we uncovered several novel symmetries of the KK
Hamiltonian for cubic $t_{2g}$ systems. It is surprising that the
KK Hamiltonian has been widely used in the study of interesting
spin-orbital physics of transition metal oxides  for a long time
but yet its remarkable symmetry properties were missed until now.
Using these symmetries, we rigorously showed that the KK
Hamiltonian without spin-orbit interactions does not permit the
development of long-range spin order in a three dimensional cubic
lattice at nonzero temperature. Inclusion of spin-orbit
interactions allows the formation of long-range spin order, but
the excitation spectrum is gapless.

ABH thanks NIST for its hospitality during several visits when
this work was done. We acknowledge partial support from the
US-Israel Binational Science Foundation (BSF). The TAU group is
also supported by the German-Israeli Foundation (GIF).




\end{document}